# Indigenization of Urban Mobility

Zimo Yang[1,2], Defu Lian[2,3,4], Nicholas Jing Yuan[2], Xing Xie[2], Yong Rui[2], Tao Zhou[1,2,4]

[1]CompleX Lab, Web Sciences Center, University of Electronic Science and Technology of China, Chengdu 611731, People's Republic of China

[2]Microsoft Research, Beijing 100080, People's Republic of China

[3]Department of Computer Science and Technology, University of Science and Technology of China, Hefei, 230026, People's Republic of China

[4]Big Data Research Center, University of Electronic Science and Technology of China, Chengdu 611731, People's Republic of China

**Abstract**

The identification of urban mobility patterns is very important for predicting and controlling spatial events. In this study, we analyzed millions of geographical check-ins crawled from a leading Chinese location-based social networking service (Jiepang.com), which contains demographic information that facilitates group-specific studies. We determined the distinct mobility patterns of natives and non-natives in all five large cities that we considered. We used a mixed method to assign different algorithms to natives and non-natives, which greatly improved the accuracy of location prediction compared with the basic algorithms. We also propose so-called indigenization coefficients to quantify the extent to which an individual behaves like a native, which depends only on their check-in behavior, rather than requiring demographic information. Surprisingly, the hybrid algorithm weighted using the indigenization coefficients outperformed a mixed algorithm that used additional demographic information, suggesting the advantage of behavioral data in characterizing individual mobility compared with the demographic information. The present location prediction algorithms can find applications in urban planning, traffic forecasting, mobile recommendation, and so on.

## 1. Introduction

Understanding urban mobility of human is very important for disease control [1], city planning [2], and traffic forecasting [3], as well as for increasing business value in location-based services [4,5]. Individuals differ greatly in their mobility patterns [6], but aggregated analyses detect regular displacement distributions, which range from power laws [7,8] to exponential laws [9,10]. These statistical regularities may be the result of combining several parameters, including the preferential return mechanism embedded in individual behaviors [11], the structure of transportation systems [12,13], urban organization [14] and the constraints of travel costs [6,15].

In general, an in-depth understanding can be obtained by classifying people into groups according to their demographic features. For some specific measures, such as predictability, group-based differences are statistically insignificant [16], but recent empirical studies have shown that people with different jobs [6], different ages [17], different genders [17–19] and different purposes (tour or not) [20] have different mobility patterns. Thus, group-based analyses are helpful when addressing specific social problems, e.g., gender-specific mobility patterns can be used to quantify the equality between men and women [20,21], the leisure mobility of elderly people can be considered as an important indicator that characterizes their living conditions [22], and the urban mobility of children may affect their future integration into society [23].

These group-based analyses can be improved in two ways. First, the demographic information contains noise and bias, where the former is derived from false or out-of-date data whereas the latter is a result of inconsistencies between demographic features and behavioral patterns; thus, at the individual level, a girl could behave like a boy and an aged person could behave like a young man. Methodologically, we can uncover the quantitative differences in behavioral patterns among people in different demographic groups, and then consider the behavioral differences directly instead of demographic differences. This shift from demographic analyses to behavioral analyses can extend the potential applications to datasets without demographic information as well as facilitating an associated shift from group-based services to personalized services. Second, group-specific insights can be used to address relevant problems such as location prediction and travel recommendation, where the algorithmic performance verifies the significance of the analyses.

In this study, we investigated the different mobility patterns of natives and non-natives, which are increasingly important in terms of business value in location-based services [24] and social value when measuring the social integration of immigrants during globalization and urbanization processes [25,26]. Our study involved the intensive analysis of a large-scale dataset, which included millions of geographical check-ins in five large cities in China. The study makes three main contributions, as follows. (i) We identified the distinct mobility patterns of natives and non-natives, i.e., the distribution of the location visiting frequencies of non-natives was more heterogeneous than that of natives at the aggregated level, whereas the frequency distributions of natives were statistically more heterogeneous at the individual level. (ii) Compared with the basic algorithms, the accuracy of location prediction was improved greatly by assigning different algorithms to natives and non-natives. (iii) We developed behavior-based indices to characterize how an individual behaves like a native, i.e., indigenization coefficients, and we showed that a hybrid algorithm weighted using indigenization coefficients, which did not require any demographic information, improved the prediction accuracy greatly.

Our study is closely relevant to the real world, in particular, the prediction of locations can find wide applications. Years ago, it is already known as a critical part in traffic forecasting [3] and urban planning [27]. Very recently, thanks to the development of information and communication technologies such as mobile communication, the mobility prediction plays an increasingly important role in location-based recommendation [4-5,28-30]. In addition, the prediction methods can be improved by considering social ties between target users while such methods can also be applied in analyzing social networks [31-35].

## 2. Data

Our experiments were based on check-in records crawled from *Jiepang* (http://jiepang.com/), which is a leading Chinese location-based social networking service that is similar to *Foursquare* (https://foursquare.com/). It helps users to record and track all of their life activities, to connect with friends at specific moments, and to explore communities of people with similar interests. Tweets attaching the specific webpage URLs of *Jiepang* check-in locations are often shared on the *weibo* platform (http://weibo.com/). We crawled these tweets via weibo public APIs and extracted location check-in information from them, including locations, time-stamps and optional texts. Notice that, all data were obtained via the open APIs. Therefore the use of these data has obeyed the corresponding terms and conditions. The data set covers the activities of users from 5 August 2011 to 17 September 2012. The detailed techniques in crawling and cleaning the data can be found in [36-37]. The data were anonymized before this study, where both user identities and location entities are replaced by 128-bit MD5 numbers, and the full dataset is available from the link (we will make it available to the public after acceptance).

**Table 1**: Basic statistics for the data used in this study. For each of the five cities, $N^u$, $N^u_y$, $N^u_n$, $N^l$, $N^l_y$, $N^l_n$, $N^c$, $N^c_y$, and $N^c_n$ represent the number of users, native users, non-native users, visited locations, locations visited by natives, locations visited by non-natives, check-ins, native check-ins, and non-native check-ins, respectively.

| City | $N^u$ | $N^u_y$ | $N^u_n$ | $N^l$ | $N^l_y$ | $N^l_n$ | $N^c$ | $N^c_y$ | $N^c_n$ |
|---|---|---|---|---|---|---|---|---|---|
| Beijing | 11077 | 6824 | 4253 | 28100 | 26882 | 8617 | 511133 | 384222 | 126911 |
| Shanghai | 6322 | 3847 | 2475 | 45070 | 44413 | 6594 | 782677 | 734719 | 48958 |
| Nanjing | 2132 | 177 | 1955 | 2421 | 1539 | 1747 | 18757 | 7934 | 10823 |
| Chengdu | 2320 | 173 | 2147 | 2330 | 1518 | 1542 | 21952 | 7976 | 13976 |
| Hong Kong | 1727 | 130 | 2597 | 2460 | 1036 | 2079 | 36775 | 5124 | 31651 |

In the dataset, each item (i.e., a check-in) recorded the user ID (anonymized), check-in time, longitude, latitude, and name of a location. The latitude, longitude and name together comprise a specific and unique location, and thus determine its representation MD5. Those mentioned frequencies are computed based on the representation MD5s of locations. Given a city *X*, all of the users were divided into three classes: (A) users whose hometown and most frequently visited city was both *X*; (B) users whose hometowns were not *X* and their most frequently visited cities were not *X*; (C) users without hometown information or who did not belong to (A) or (B). The users in classes *A* and *B* were called natives and non-natives of city *X*, respectively, but the users in (C) were not considered in this study. We used these classifications to minimize the bias caused by users who might live and work for a long time in a city that differed from their hometown.

We targeted five big cities in China: Beijing, Shanghai, Nanjing, Chengdu, and Hong Kong. Users who checked in only once and locations that appeared only once were removed. After filtering, the data comprised 1,371,294 check-ins. The basic statistics for the dataset are shown in Table 1. It should be noted that user records could appear in several places because they could be a native of Beijing but also a non-native of Chengdu and Nanjing, whereas a check-in could only occur in one place.

## 3. Empirical Analysis

As shown in Figure 1, for each city, the distribution of the location visiting frequencies for non-natives is more heterogeneous than that for natives. This is because the non-natives tended to visit popular locations, such as the Imperial Palace in Beijing and the Bund in Shanghai. In contrast, natives usually check in repeatedly at locations of personal importance, and these locations are different for different natives. Therefore, the distribution of visiting frequencies contributed by all natives is relatively homogeneous.

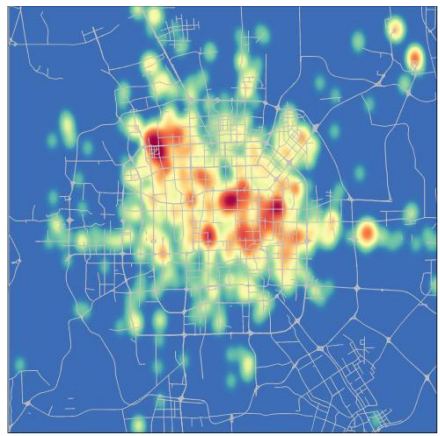

(a) Native check-ins in Beijing

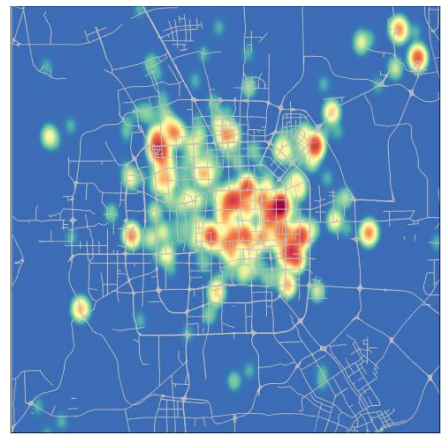

(b) Non-native check-ins in Beijing

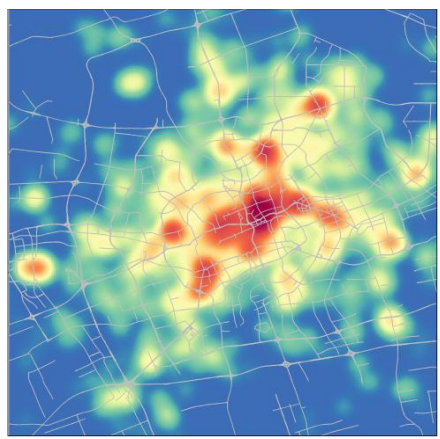

(c) Native check-ins in Shanghai

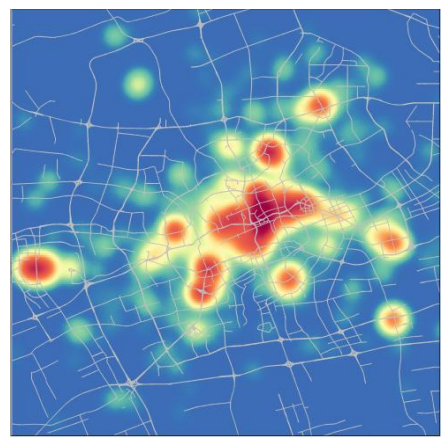

(d) Non-native check-ins in Shanghai

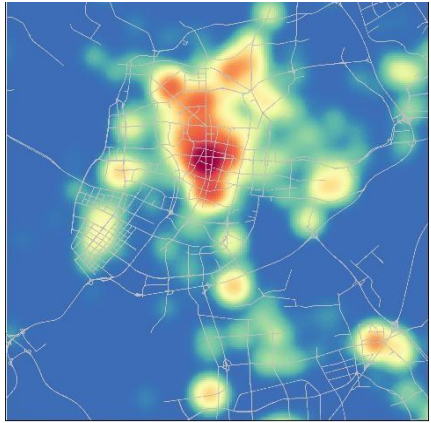
(e) Native check-ins in Nanjing

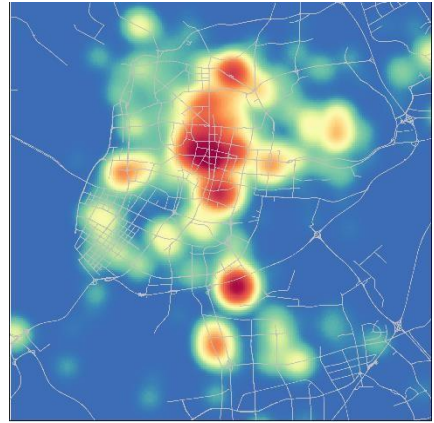
(f) Non-native check-ins in Nanjing

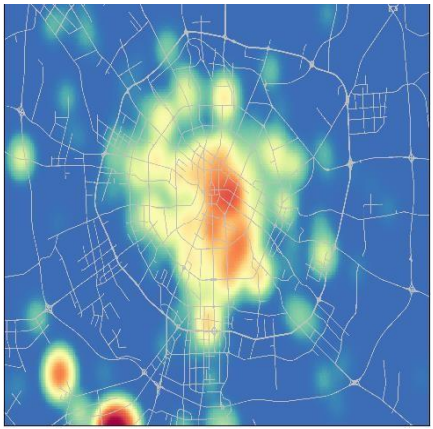
(g) Native check-ins in Chengdu

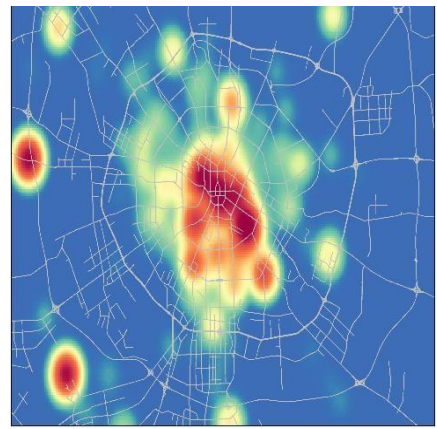
(h) Non-native check-ins in Chengdu

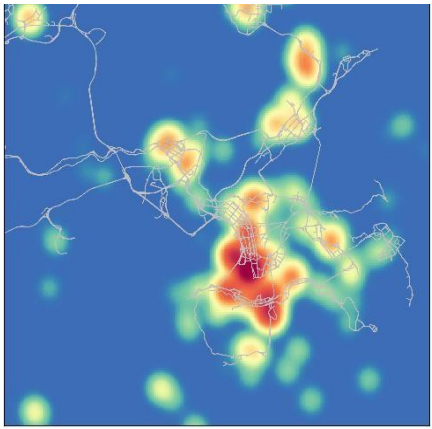
(i) Native check-ins in Hong Kong

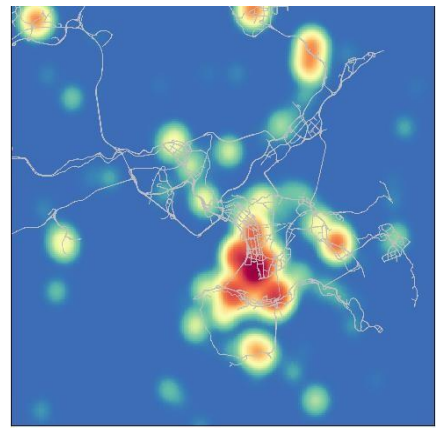
(j) Non-native check-ins in Hong Kong

**Figure 1**: Density maps of location visiting frequencies for: (a) natives in Beijing, (b) non-natives in Beijing, (c) natives in Shanghai, (d) non-natives in Shanghai, (e) natives in Nanjing, (f) non-natives in Nanjing, (g) natives in Chengdu, (h) non-natives in Chengdu, (i) natives in Hong Kong, and (j) non-natives in Hong Kong. The frequencies were normalized for each plot and the areas colored in red have high densities.

We used the well-known Gini coefficient [38] to quantify the heterogeneity of a visiting frequency distribution. The Gini coefficient measures the inequality among the values in a frequency distribution, which can theoretically range from 0 to 1, where 0 and 1 correspond to complete equality and inequality, respectively. A higher Gini coefficient indicates a more heterogeneous distribution, and vice versa.

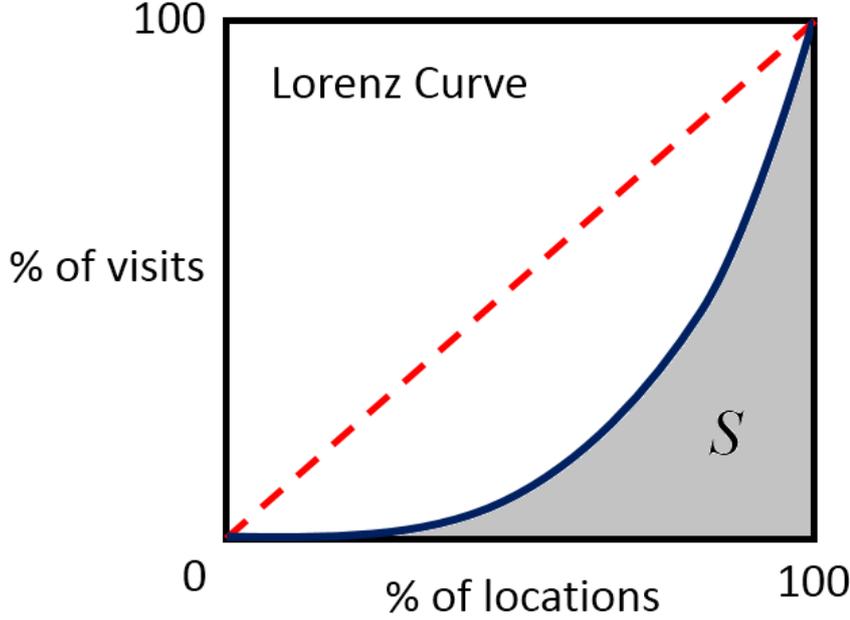

**Figure 2**: An illustration of the Lorenz curve. The red dash line stands for the ideal case that every location has been visited exactly the same times, and the blue curve is an example Lorenz curve, with the shadowed area being *S*.

The Gini coefficient is usually defined mathematically based on the Lorenz curve, which is used in characterizing the heterogeneity of a distribution. For example, in this paper, when drawing a Lorenz curve for the distribution of locations' visits (which can be contributed by all visitors, only native visitors, or only non-native visitors), we firstly rank all locations according to their visiting frequencies. As shown in figure 2, the x-axis stands for the fraction of locations where all locations are arranged from the less visited ones (left) to frequently visited ones (right), and the y-axis represents the fraction of cumulated number of visits on those locations. If every location has been visited exactly the same times, the Lorenz curve is the diagonal line as the red dash line in figure 2. The blue line is a usual example of Lorenz curve in real heterogeneous systems. For example, a point (0.4, 0.05) in the blue curve means the 40% of the least visited locations contribute only 5% of the total visits. Obviously, for a highly heterogeneous distribution, the area S under the Lorenz curve is very small, therefore we adopt the simple coefficient G=1-2S to quantify the extent of heterogeneity, which is the well-known Gini coefficient. If there are *n* locations with visiting frequencies of $f_1<=f_2<=...<=f_n$, the Gini coefficient is:

$$G = \frac{1}{n}\left\{n+1-2\left[\frac{\sum_{i=1}^{n}(n+1-i)f_i}{\sum_{i=1}^{n}f_i}\right]\right\} = \frac{2\sum_{i=1}^{n}if_i}{n\sum_{i=1}^{n}f_i} - \frac{n+1}{n}. \qquad (1)$$

**Table 2**: Gini coefficients for the distributions of visiting frequencies at different locations. $G_y^p$ and $G_n^p$ denote the coefficients contributed by the native population and non-native population, respectively, while $\langle G_y^i \rangle$ and $\langle G_n^i \rangle$ denote the average coefficients for native individuals and non-native individuals.

| City | $G_y^p$ | $G_n^p$ | $\langle G_y^i \rangle$ | $\langle G_n^i \rangle$ |
|---|---|---|---|---|
| **Beijing** | 0.68 | 0.77 | 0.21 | 0.14 |
| **Shanghai** | 0.69 | 0.83 | 0.26 | 0.13 |
| **Nanjing** | 0.43 | 0.68 | 0.24 | 0.15 |
| **Chengdu** | 0.43 | 0.77 | 0.24 | 0.10 |
| **Hong Kong** | 0.46 | 0.82 | 0.21 | 0.16 |

As shown in Table 2, for every city that we considered, at the aggregated level, the Gini coefficient contributed by non-natives was higher than the Gini coefficient contributed by natives (i.e., $G_n^p > G_y^p$), which is in accordance with Fig. 1. The corresponding Lorenz curves are presented in Fig. 3.

At the individual level, a native individual tends to check in repeatedly at locations of personal importance, while a non-native individual does not stay long in an area and thus they did not check in multiple times at specific locations. Therefore, the visiting frequency distribution of a native person was significantly more heterogeneous than that of a non-native person. As shown in Table 1, this hypothesis was supported because the average Gini coefficient for all native individuals was higher than that for all non-natives, i.e., $\langle G_y^i \rangle > \langle G_n^i \rangle$, for every city under consideration. The significance of the two results, $G_n^p > G_y^p$ and $\langle G_y^i \rangle > \langle G_n^i \rangle$, has been validated by well-accepted statistical tests including t-statistics, degree of freedom and p-value [39].

Natives and non-natives are also statistically distinguishable in other aspects. As shown in Figure 2(a)-(d), the native individuals statistically have visited more distinct locations (i.e., larger $N_D$ in average) and shared more check-ins (i.e., larger $N_T$ in average). As shown in Figure 2(e) and Figure 2(f), the natives and non-natives have remarkable differences in their active periods, where an individual's active period is simply quantified by the number of days between his first and last check-ins. Firstly, a native user's active period is much longer than a non-native user, with a significant peak around 270 days for all five cities. Secondly, a small but notable peak at about a week is observed in each distribution for non-native users, which is probably contributed by the tourists.

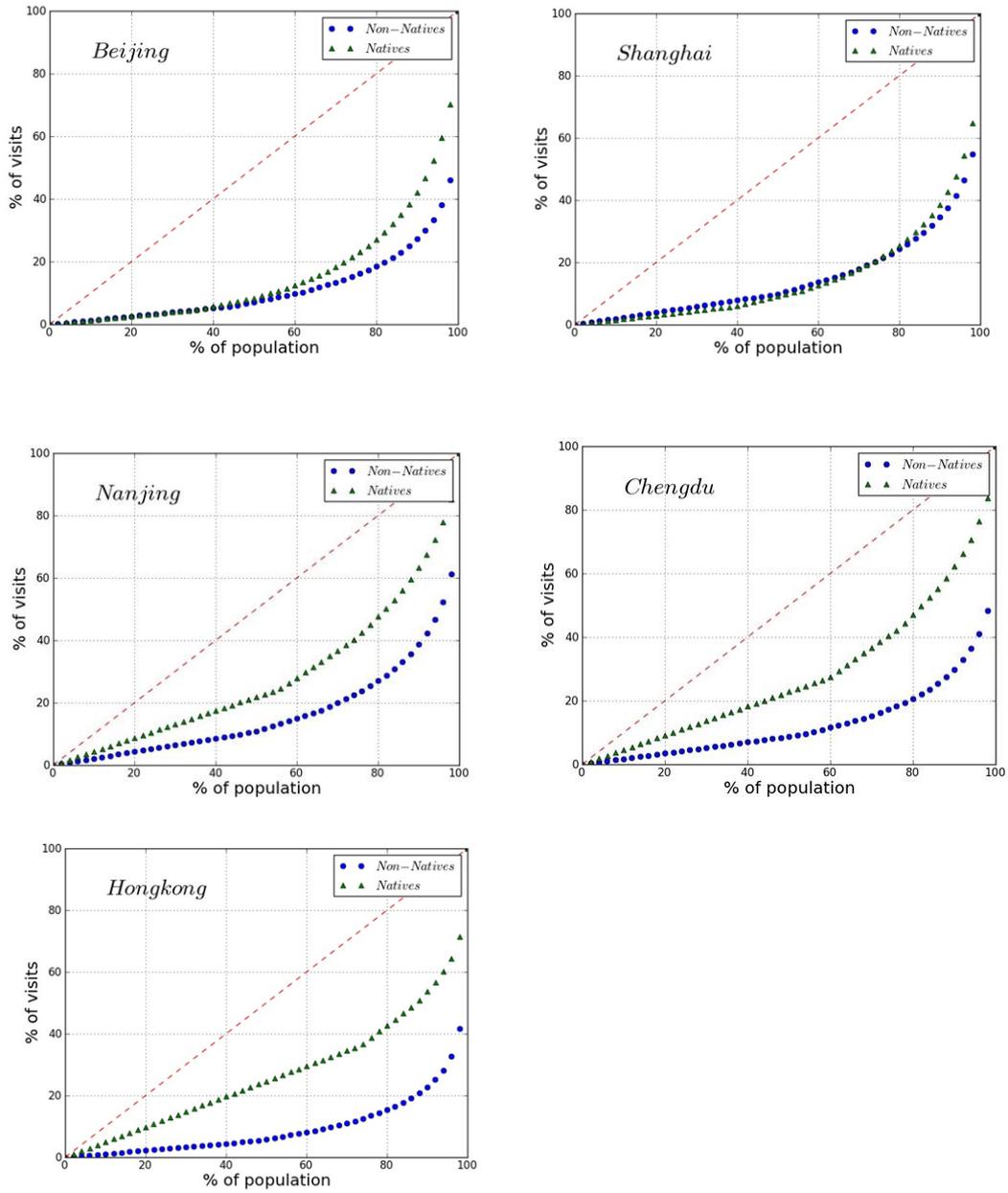

**Figure 3**: Lorenz curves at the population level for Beijing, Shanghai, Nanjing, Chengdu and Hong Kong. The triangles and circles represent the results for natives and non-natives, respectively. In each city, the distribution of visited frequencies of locations that contributed by all non-natives is more heterogeneous than that contributed by all natives.

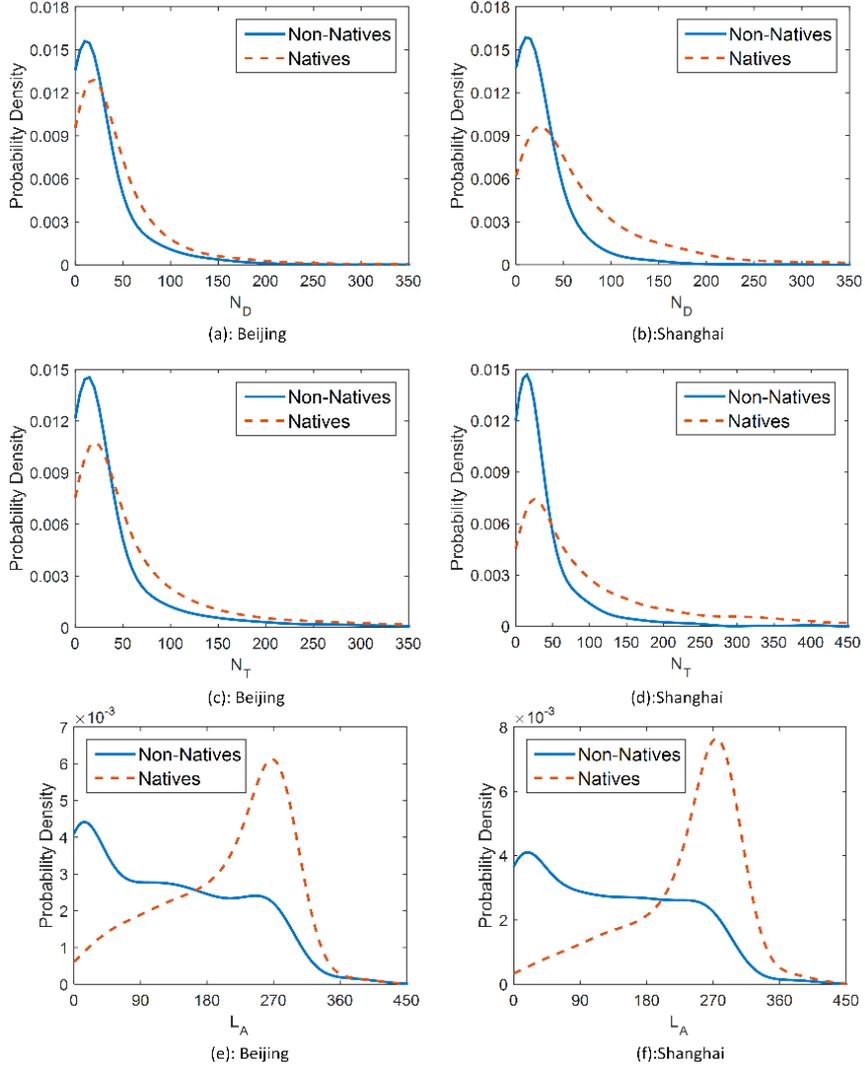

**Figure 4**: Statistical differences between natives and non-natives. The red solid and blue dash curves stand for probability density functions for natives and non-natives, respectively. The six subplots respectively report the distributions of the number of distinct visited locations $N_D$ in (a) Beijing and (b) Shanghai, the number of total check-ins $N_T$ in (c) Beijing and (d) Shanghai, the length of active period $L_A$, namely the number of days between a user's first and last check-ins in (e) Beijing and (f) Shanghai. These distributions for the other three cities (i.e., Nanjing, Chengdu and Hong Kong) are similar and thus omitted.

## 4. Location Prediction Algorithms

Location prediction is a core technique that underlies many significant location-based services [23] and other application [40]. Thus, many algorithms have been proposed to obtain highly accurate predictions [41], including collaborative analysis, Markov chain methods, linear regression, decision trees (e.g., M5 tree and T-pattern tree), neural networks, Bayesian networks, and other data mining approaches [4,5,42-49]. In our analysis, we used the two simplest methods for location prediction. The first is called the history-based method (HB), where given a target user, each location is scored directly as the number of check-ins it receives in the target user's records. The second is called the popularity-based method (PB), where the prediction score for a location is the total number of check-ins it receives in all of the users' records, regardless of how many times it appears in the target user's check-in list. A location with a higher score is assumed to be more likely to appear in the target user's future trajectory.

The check-ins were divided into two datasets: the training set contained 90% of the check-ins and the testing set comprised the remaining 10%. Such division can be in two ways: *random division* with the check-ins in testing set being selected randomly and *temporal division* where the testing set contains the 10% newest check-in records. Every record in the testing set satisfied two conditions: (i) the user had at least two check-ins in the target city; (ii) the location was associated with at least two check-ins. Given a target user, a location prediction algorithm produces a ranked list of all locations, where those with high likelihoods of being visited by the target user occupy the top positions. It should be noted that this differs slightly from an e-commerce recommender system [50], where an item bought by the target user typically cannot be recommended again.

We evaluated the prediction accuracy using the AUC [51], namely the area under the receiver operating characteristics curve, which is similar to the ranking score [52]. The AUC measured how well a prediction algorithm could successfully distinguish relevant locations (those in the testing set) from the irrelevant locations. For a check-in ($u,\alpha$) in the testing set, if the location $\alpha$ was ranked in the $k$th place among all $N$ locations in the training set for the user $u$, its AUC value was ($N$-$k$+1)/$N$. The AUC of an algorithm was defined as the average AUC value over all check-ins in the testing set. Clearly, for a random prediction, AUC = 0.5, and for a prefect prediction, AUC = 1. Therefore, the extent to which AUC exceeded 0.5 indicated the prediction accuracy. In addition to the AUC value, we also adopt the Recall value [53]. For every user, we provide a list of $L$ predictions according to the algorithm. If a check-in ($u,\alpha$) in the testing set was ranked within the top-$L$ place, it is called correctly predicted. The Recall value is defined as the ratio of correctly predicted check-ins to the total number of check-ins. In this paper, we only show the result at $L$=100 (Recall@100). The recall values for other $L$ and the Precision value [53] will give qualitatively the same results.

Table 3 and Table 4 show the AUC values obtained using HB and PB. At the individual level, the visiting distribution of a native is statistically more heterogeneous than that of a non-native, so it is reasonable to expect that HB is more suitable for natives than non-natives. While at the population level, the visiting distribution contributed by all non-natives is more heterogeneous than that contributed by all natives, indicating that non-natives tend to visit some popular spots in a city and thus PB is more suitable for non-natives than natives. These inferences were supported by the empirical analysis. As shown in Table 5, the average prediction accuracy for natives was higher (or lower) than that for non-natives in every city under HB (or PB). Thus, we developed the so-called demography-based method (DB), which is a mixed algorithm that applies HB if the target user is a native person, or PB for a non-native person. As shown in Table 3 and Table 4, the mixed algorithm outperformed both HB and PB.

**Table 3**: Accuracy of location prediction quantified by AUC values. The six prediction algorithms presented here are history-based method (HB), popularity-based method (PB), demography-based method (DB), individual behavior-based method (IBB), collaborative-behavior-based (CBB) and logistic-regression-based method (LRB). All data points were obtained by averaging over 10 independent runs with random divisions into training and testing sets.

| City | HB | PB | DB | IBB | CBB | LRB |
|---|---|---|---|---|---|---|
| **Beijing** | 0.7830 | 0.8282 | 0.8307 | 0.9084 | 0.9077 | 0.9094 |
| **Shanghai** | 0.8462 | 0.8327 | 0.8580 | 0.9247 | 0.9234 | 0.9277 |
| **Nanjing** | 0.7855 | 0.7712 | 0.8193 | 0.8731 | 0.8638 | 0.8751 |
| **Chengdu** | 0.7914 | 0.8288 | 0.8601 | 0.8990 | 0.8921 | 0.9070 |
| **Hong Kong** | 0.7598 | 0.8875 | 0.8954 | 0.9262 | 0.9238 | 0.9292 |

**Table 4**: Accuracy of location prediction quantified by AUC values. The six prediction algorithms presented here are history-based method (HB), popularity-based method (PB), demography-based method (DB), individual behavior-based method (IBB), collaborative-behavior-based (CBB) and logistic-regression-based method (LRB). All data points were obtained by averaging over 10 independent runs with temporal divisions into training and testing sets.

| City | HB | PB | DB | IBB | CBB | LRB |
|---|---|---|---|---|---|---|
| **Beijing** | 0.7220 | 0.8286 | 0.8012 | 0.8928 | 0.8903 | 0.8953 |
| **Shanghai** | 0.8089 | 0.8321 | 0.8251 | 0.9094 | 0.9041 | 0.9102 |
| **Nanjing** | 0.7648 | 0.7692 | 0.8113 | 0.8583 | 0.8345 | 0.8645 |
| **Chengdu** | 0.7331 | 0.8229 | 0.8516 | 0.8911 | 0.8733 | 0.9054 |
| **Hong Kong** | 0.6745 | 0.9077 | 0.9022 | 0.9277 | 0.9267 | 0.9276 |

**Table 5**: AUC values for natives and non-natives under HB and PB methods. For each city, HB outperforms PB for natives while PB outperforms HB for non-natives.

| | HB | PB |
|---|---|---|
| **Beijing Natives** | 0.8575 | 0.8272 |
| **Shanghai Natives** | 0.8565 | 0.7936 |
| **Nanjing Natives** | 0.8293 | 0.7406 |
| **Chengdu Natives** | 0.8486 | 0.8014 |
| **Hong Kong Natives** | 0.8665 | 0.8005 |
| **Beijing Non-Natives** | 0.7644 | 0.8923 |
| **Shanghai Non-Natives** | 0.8018 | 0.8569 |
| **Nanjing Non-Natives** | 0.6981 | 0.8316 |
| **Chengdu Non-Natives** | 0.7129 | 0.8790 |
| **Hong Kong Non-Natives** | 0.7417 | 0.9159 |

Despite its considerable improvement, the demography-based method has two limitations. First, truthful demographic information is not easy to obtain in general online systems, and thus the applicability of the method is restricted. Second, some demographic evidence may be inconsistent with the behavioral patterns, where we aim to predict the latter, i.e., we want to quantify the extent to which an individual behaves like a native. Given that a native is more likely to visit some locations more times compared with a non-native (see the average Gini coefficients in Table 2), we propose an index $I_i$ to count the ratio of repeated check-ins, i.e., for a user $u$, $N_T^{(u)}$ denotes the total number of $u$'s check-ins and $N_D^{(u)}$ is the number of different locations visited by $u$; thus, the index is defined as:

$$I_i(u) = 1 - \frac{N_D^{(u)}}{N_T^{(u)}}. \quad (2)$$

For example, if the user $u$ has seven check-ins at locations {A,B,C,A,B,A,D}, then $N_D^{(u)} = 4$ and thus $I_i(u)$=3/7. This is an individual behavioral index because it only requires the behavioral information for an individual. Analogously, given that a native is less likely to visit popular locations than a non-native (see the Gini coefficients in Table 2), we propose an index $I_c$ to count the average normalized popularity of a user's visits. First, we rank all of the locations in descending order according to their visiting frequencies (locations with the same frequency are ranked randomly). Next, a location's normalized popularity is defined by its ranking score, e.g., if the Great Wall is ranked fifth among all 28,100 locations in Beijing, its normalized popularity is 5/28100. Thus, the $I_c$ of a user is the average normalized popularity of the user's visits, where a location that appears several times should also be counted several times. If a user $u$'s sequential check-in locations are $l_1, l_2, \cdots, l_{N_T^{(u)}}$, then the index is:

$$I_c(u) = \frac{1}{N_T^{(u)}} \sum_{k=1}^{N_T^{(u)}} R(l_k), \quad (3)$$

where $R(l_k)$ is the normalized popularity (i.e., ranking score) of location $l_k$. In contrast to $I_i$, $I_c$ is a collaborative behavioral index because it requires the behavioral information for all users. $I_i$ and $I_c$ are called indigenization coefficients, and a larger value indicates a higher similarity to a native for both coefficients.

Thus, we propose an individual-behavior-based (IBB) method. Given a target user $u$, the predicted score for a location $l$ is:

$$S_i(u,l) = I_i^\alpha(u) P(u,l) + [1 - I_i(u)]^\alpha Q(l), \quad (4)$$

where $\alpha$ is a free parameter, $P$ is the normalized history score for user $u$ and location $l$, which is defined as the ratio of the number of $u$'s check-ins at location $l$ relative to the total number of $u$'s check-ins, and $Q$ is the normalized popularity score for location $l$, which is defined as:

$$Q(l) = N(l) \big/ \max_{l'} N(l'), \quad (5)$$

where $N(l)$ is the number of visits at location $l$ by all users. Analogously, a collaborative-behavior-based (CBB) method scores a location $l$ as follows:

$$S_c(u,l) = I_c^\alpha(u) P(u,l) + [1 - I_c(u)]^\alpha Q(l). \quad (6)$$

Table 3 and Table 4 shows the AUC values for IBB and CBB with the optimal parameter $\alpha$. Surprisingly, without using any demographic information, the accuracies of both IBB and CBB were improved greatly compared with the demography-based method.

**Table 6**: Accuracy of location prediction quantified by Recall with L=100. The six prediction algorithms presented here are history-based method (HB), popularity-based method (PB), demography-based method (DB), individual behavior-based method (IBB), collaborative-behavior-based (CBB) and logistic-regression-based method (LRB). All data points were obtained by averaging over 10 independent runs with random divisions into training and testing sets.

| Recall | HB | PB | DB | IBB | CBB | LRB |
|---|---|---|---|---|---|---|
| **Beijing** | 0.5409 | 0.2247 | 0.5507 | 0.5974 | 0.5774 | 0.6054 |
| **Shanghai** | 0.6411 | 0.1938 | 0.6436 | 0.6639 | 0.6379 | 0.6781 |
| **Nanjing** | 0.6275 | 0.5712 | 0.6873 | 0.7849 | 0.7564 | 0.7957 |
| **Chengdu** | 0.6461 | 0.7151 | 0.7682 | 0.8383 | 0.8315 | 0.8396 |
| **Hong Kong** | 0.5669 | 0.8062 | 0.8248 | 0.8885 | 0.8851 | 0.8902 |

**Table 7**: Accuracy of location prediction quantified by Recall with L=100. The six prediction algorithms presented here are history-based method (HB), popularity-based method (PB), demography-based method (DB), individual behavior-based method (IBB), collaborative-behavior-based (CBB) and logistic-regression-based method (LRB). All data points were obtained by averaging over 10 independent runs with temporal divisions into training and testing sets.

| Recall | HB | PB | DB | IBB | CBB | LRB |
|---|---|---|---|---|---|---|
| **Beijing** | 0.4253 | 0.2337 | 0.5061 | 0.5049 | 0.5207 | 0.5271 |
| **Shanghai** | 0.5746 | 0.2142 | 0.5899 | 0.6098 | 0.5998 | 0.6134 |
| **Nanjing** | 0.5912 | 0.5776 | 0.6781 | 0.7536 | 0.7386 | 0.7612 |
| **Chengdu** | 0.5387 | 0.684 | 0.7513 | 0.8248 | 0.8084 | 0.8319 |
| **Hong Kong** | 0.4788 | 0.8408 | 0.8425 | 0.8957 | 0.8917 | 0.8968 |

Table 6 and Table 7 show the recall values at *L*=100 for all the above-mentioned algorithms, which are in accordance with the results of AUC values.

## 5. Discussion

Based on the distinct mobility patterns of natives and non-natives, we developed a mixed algorithm that uses demographic information, which greatly outperformed basic algorithms based on the visiting history of individuals and location popularity. We also developed two indigenization coefficients, $I_i$ and $I_c$, for estimating the extent to which an individual behaves like a native. Without any demographic information, the simple hybrid algorithm weighted using each indigenization coefficient greatly improved the prediction accuracy compared with the mixed algorithm (DB). It is not surprising that grouping people before applying mixed or hybrid algorithms (e.g., see [54]) can yield more accurate predictions than those obtained with the basic algorithms. However, we must emphasize that it is not necessary to group people based on their attributes, but instead we can simply characterize them based on behavioral coefficients after determining the correlations between attributes and behavioral patterns. Shifting from demographic information to indigenization coefficients is only an example; indeed, we could build a user's profile based only on their behavioral records and select effective features based on an in-depth understanding of the relationships between attributes and behaviors. This method differs from attribute-based methods in two respects: (i) it can be applied without attribute information because knowledge of how the behavioral features are selected can be learned from other similar systems with attribute information [55]; (ii) if the selections are appropriate, a number of quantitative behavioral coefficients can provide a more accurate description of users than a static classification based on attributes. Furthermore, this method differs from mainstream machine learning methods (e.g., the ensemble learning method [56]) because it can provide insights in addition to predictions.

Although the current prediction accuracies are competitive (for example, even HB can provide slightly more accurate predictions than the one-order Markov chain method [43] that gives AUC values as 0.7819, 0.8460, 0.7723, 0.7858 and 0.7588 for Beijing, Shanghai, Nanjing, Chengdu and Hong Kong, respectively), the readers should be aware that this paper does not aim at the design of a better-performed yet complicated algorithm than other state-of-the-art ones, but to reveal the difference between mobility patterns of natives and non-natives, as well as to show the power of behavioral data analysis. It is very possible that a machine learning algorithm with carefully selected group of features via the feature engineering techniques can beat the current algorithm in prediction accuracy. Therefore, we emphasize phenomena and perspectives, rather than the details and refinements of algorithms. Indeed, our proposed prediction algorithms can be improved further in many ways. For example, the prediction accuracy of IBB and CBB can be improved further by introducing a stretched index to make the mean values of $P(u,l)$ and $Q(l)$ the same, although this would make the equation and learning process more complicated. We can also define an integrated indigenization coefficient:

$$I = \frac{1}{1+\exp(-w_i I_i - w_c I_c)}, \qquad (7)$$

where the parameters $w_i$ and $w_c$ can be learned from the logistic regression that best classifies natives and non-natives. As shown in Table 3, Table 4, Table 6 and Table 7, this logistic-regression-based (LRB) method can improve the accuracy further, but it requires the demographic information and the learned parameters for different cities are very different.

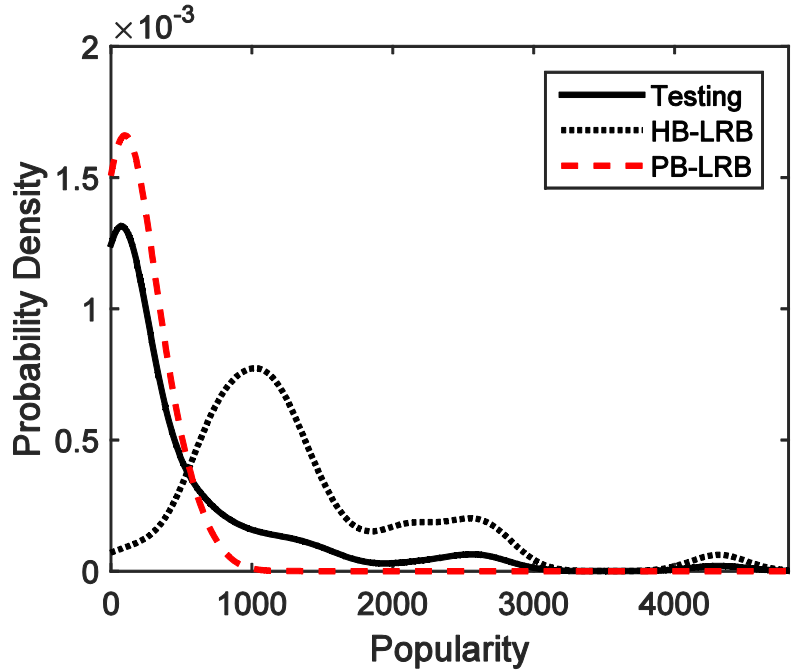

**Figure 5**: The probability density functions of visiting frequencies of locations (i) in the testing set (back solid curve), (ii) having not been correctly predicted by HB while been correctly predicted by LRB (red dash curve), and (iii) having not been correctly predicted by PB while been correctly predicted by LRB (black dot curve). If a location has been appeared several times in different check-ins, it will be counted several times. This figure shows the result for Beijing, and the other four cities are similar.

We have recorded all the check-ins in the testing set which have not been correctly predicted by HB or PB, but been correctly predicted by LRB. These two sets are respectively denoted by HB->LRB and PB->LRB. Figure 5 shows the distributions of visiting frequencies of corresponding locations in HB->LRB and PB->LRB for Beijing (other cities are similar). For PB->LRB, the result is trivial since only less attractive locations can be missed in PB, namely in the set PB->LRB, there are no popular locations. For HB->LRB, the distribution is remarkably boarder than that of all corresponding locations in the testing set, indicating that individuals also prefer to visit popular yet unvisited locations, which is a typical behavioral pattern of non-natives. Therefore, the LRB algorithm can be to some extent treated as a tradeoff of HB and PB. To compare HB or PB with IBB or CBB will achieve the similar results.

As possibly future works, if we have sufficiently long-term and large-scale datasets, we can measure the trend in the integrated indigenization coefficient and estimate whether a non-native person will behave like a native person after a specific time period and, if this is the case, how long the average person requires to act like a native. At the same time, taking into account the temporal information, one may design a time-aware indigenization coefficient that can more effectively distinguishing different behavioral patterns between natives and non-natives and lead to more accurate location prediction. Besides, the cross-region (i.e., data from different cities in different countries) and cross-data-type (i.e., GPS, RFID, Wi-Fi and other types of data sets) analyses are very helpful in validating the current findings and uncovering other mobility patterns.

## Acknowledgments

The authors acknowledge their helpful discussions with Xiaoyong Yan. This research was supported partly by the National Natural Science Foundation of China under Grant Nos. 11222543 and 61433014, the Program for New Century Excellent Talents in University under Grant No. NCET-11-0070, and the CCF-Tencent open fund.